\documentclass[prl,twocolumn,superscriptaddress,nofootinbib]{revtex4}

\usepackage{amsfonts}
\usepackage{amsmath}
\usepackage{amssymb}
\usepackage{amsthm}
\usepackage{bm}
\usepackage{dcolumn}
\usepackage{epsfig}
\usepackage{graphicx}
\usepackage{graphics}
\usepackage[latin1]{inputenc}
\usepackage{latexsym}
\usepackage{rotating}
\usepackage{hyperref}

\newcommand\be{\begin{equation}}
\newcommand\ba{\begin{eqnarray}}
\newcommand\ee{\end{equation}}
\newcommand\ea{\end{eqnarray}}

\newcommand{\pont}{{\,^\ast\!}R\,R}

\begin{document}
\title{Linear Stability Analysis and the Speed of Gravitational Waves \\ 
in Dynamical Chern-Simons Modified Gravity}

\author{David Garfinkle}
\affiliation{Department of Physics, Oakland University, Rochester, MI 48309, USA,
and Michigan Center for Theoretical Physics, Randall Laboratory of Physics, University of 
Michigan, Ann Arbor, MI 48109-1120, USA.}

\author{Frans Pretorius}
\affiliation{Department of Physics, Princeton University, Princeton, NJ 08544, USA.}

\author{Nicol\'as Yunes}
\affiliation{Department of Physics, Princeton University, Princeton, NJ 08544, USA.}

\date{\today}

\begin{abstract}

We perform a linear stability analysis of dynamical Chern-Simons modified gravity 
in the geometric optics approximation and find that it is
linearly stable on the backgrounds considered. Our analysis also reveals that 
gravitational waves in the modified theory travel at the speed of light in Minkowski spacetime.
However, on a Schwarzschild background the characteristic speed of propagation along
a given direction splits into two modes, one subluminal and one superluminal.
The width of the splitting depends on the azimuthal components of the propagation
vector, is linearly proportional to the mass of the black hole, and decreases
with the third inverse power of the distance from the black hole.
Radial propagation is unaffected, implying that as probed
by gravitational waves the location of the event horizon of the spacetime
is unaltered. The analysis further reveals that when a high frequency, pure
gravitational wave is scattered from a black hole, a scalar wave of 
comparable amplitude is excited, and vice-versa.

\end{abstract}

\pacs{04.50.Kd,04.30.-w04.30.Nk,,04.25.-g}




\maketitle

{\emph{Introduction}}.
The unknown nature of dark matter and dark energy, and the possibility that
one or both of these phenomena could be a misinterpretation of observations
if general relativity is not the correct theory of spacetime on all
scales, has renewed interest in alternative theories of gravity.
One particular alternative theory that has attracted considerable attention recently is dynamical 
Chern-Simons (CS) modified gravity~\cite{Alexander:2009tp}. This theory is interesting because it passes 
all Solar System and binary pulsar constraints, while allowing for large deviations in highly curved
spacetimes, such as for example around rotating black holes~\cite{Yunes:2009hc,Konno:2009kg}. Moreover, 
terms akin to the CS modification in the action generically arise 
in string theory~\cite{Alexander:2004xd,Gates:2009pt}, effective field theory~\cite{Weinberg:2008hq} 
and also possibly in loop quantum gravity~\cite{Taveras:2008yf,Mercuri:2009zt}. 
In this sense, CS modified gravity can be considered theoretically well-motivated. The theory is also
an excellent proxy for the study of gravitational parity violation, as it includes the only
non-vanishing curvature-squared term that violates this symmetry~\cite{Alexander:2009tp}.

Chern-Simons modified gravity extends the Einstein-Hilbert action through the addition of the product of a 
scalar field with the Pontryagin invariant, which is the contraction of the Riemann tensor and its dual. In the dynamical
theory studied here, the action also includes a kinetic term for the scalar field. The Pontryagin
invariant is a curvature squared correction that breaks parity and time-reversal invariance, 
due to the Levi-Civita tensor in the definition of the Riemann dual. Such a term leads to modifications
to the field equations that alter General Relativity (GR) solutions that break parity. 
Examples of such GR solutions are
the Kerr spacetime or a binary compact object spacetime, where either the spin or the total angular
momentum vectors select a preferred direction that couples to the CS correction. Thus, for example, though
Schwarzschild is a solution to CS modified gravity, 
Kerr is not.

The analytical complexity of the CS modified field equations makes it almost impossible to find 
closed-form solutions to astrophysically interesting spacetimes. An alternative is to evolve such 
spacetimes numerically. However, the modified field equations generically include third time derivatives in the form
of contractions of the Levi-Civita tensor with the covariant derivative of the Ricci scalar, which 
could render the theory ill-posed and unstable. In fact, Ostrogradski showed many years ago
that theories defined by Lagrangians with more than one time derivative, which cannot be eliminated
by partial integration, are linearly unstable~\cite{Ostrogradski}. If one considers
CS modified gravity to be the leading order correction to GR of some putative well
behaved gravitational theory, then it might be reasonable to expect that the
solutions of the stable sector of the effective theory are good approximations to the
physically-relevant solutions of the full theory. 
In principle, the stable sector can be found by order reduction, i.e., performing a perturbative expansion in the 
coupling parameter of the effective theory~\cite{Woodard:2006nt}.
A numerical evolution of CS modified gravity will therefore
need to implement a similar algorithm, rather than attempt a ``brute-force'' solution.

Whether dynamical CS gravity, even when considered as an effective theory, 
is mathematically well-posed is unknown at present. 
One necessary (though not sufficient) condition for well-posedness is linear stability
of the theory about backgrounds that are stable in GR. Clearly, if the background is flat,
then CS gravity is linearly stable as higher derivatives can only arise at higher than linear order, since
the Riemann tensor vanishes. Thus, a meaningful test of linear stability requires 
study of perturbations about a solution to CS modified gravity that has non-trivial curvature.

In this paper we perform a linear stability analysis in the geometric optics (high-frequency) approximation.
We derive general relationships valid for an arbitrary background geometry, then evaluate
these in a few specific spacetimes, including Minkowski and Schwarzschild. We show explicitly that dynamical CS gravity
is linearly stable on these backgrounds, though for Schwarzschild this is only true when
the sign of the coupling constant is chosen such that the CS scalar field possesses positive 
energy\footnote{Incidentally, this physically-reasonable choice is opposite to the sign used in an earlier study of quasi-normal 
modes on a Schwarzschild background~\cite{Cardoso:2009pk}, which explains why~\cite{Cardoso:2009pk}
found unstable modes}. As a byproduct of this analysis, we derive the speed of 
coupled gravitational wave/CS-scalar modes, and show that, while in Minkowski spacetime
it is exactly the speed of light, in a Schwarzschild spacetime it differs from this speed,
allowing for both subluminal and superluminal modes. One way to physically understand this is to consider 
the CS-modifications to the Einstein field equations as an additional effective stress-energy tensor for 
the gravitational field; this effective matter violates the usual energy conditions, allowing for 
superluminal modes.

In this paper we follow mostly the conventions of~\cite{Misner:1964}. We use abstract index notation,
the metric signature $(-,+,+,+)$, geometric units where $G=c=1$, and overhead bars to denote
quantities associated with the background spacetime.
 
{\emph{Basics of CS gravity}}. The CS modified action leads to the following field equations
(we refer the reader to~\cite{Alexander:2009tp} for the action and other
details):
\ba
\label{eom}
G_{ab} + \frac{\alpha}{\kappa} C_{ab} &=& \frac{1}{2 \kappa} T_{ab},
\\
\label{eq:constraint}
\beta \; \square \vartheta &=& 
- \frac{\alpha}{4} \pont,
\ea
where $G_{ab}$ is the Einstein tensor, $\square$ is the D'Alembertian operator, $\kappa = (16 \pi G)^{-1}$, 
and the C-tensor is
\be
\label{Ctensor}
C^{ab} = (\nabla_c \vartheta)
\epsilon^{cde(a}\nabla_eR^{b)}{}_d+(\nabla_{c} \nabla_{d} \vartheta) {\,^\ast\!}R^{d(ab)c}\,,
\ee
with the dual Riemann tensor defined as
\be
\label{Rdual}
{^\ast}R^a{}_b{}^{cd}=\frac12 \epsilon^{cdef}R^a{}_{bef}\,,
\ee
and $\epsilon^{cdef}$ the 4-dimensional Levi-Civita tensor. In this study
we do not consider any potential for the CS scalar (see e.g.,~\cite{Alexander:2009tp}). 
The Pontryagin invariant is defined as
\be
\label{pontryagindef}
\pont= {\,^\ast\!}R^a{}_b{}^{cd} R^b{}_{acd} = \frac12 \epsilon^{cdef}R^a{}_{bef}  R^b{}_{acd}\,,
\ee
while the total stress-energy tensor is   
\be
\label{Tab-total}
T_{ab} = T^{\textrm{mat}}_{ab} + T_{ab}^{\vartheta}, 
\ee
where $T^{\textrm{mat}}_{ab}$ is the matter stress-energy, which we set to zero here~\cite{Alexander:2009tp},
and $T_{ab}^{\vartheta}$ is the stress-energy of the CS field $\vartheta$, given by
\be
\label{Tab-theta}
T_{ab}^{\vartheta} 
=   \beta  \left[  \left(\nabla_{a} \vartheta\right) \left(\nabla_{b} \vartheta\right) 
    - \frac{1}{2}  g_{a b}\left(\nabla_{c} \vartheta\right) \left(\nabla^{c} \vartheta\right) 
\right].
\ee
The quantities $(\alpha,\beta)$ are coupling constants whose dimensions depend on the dimensions
one chooses for the scalar field~\cite{Yunes:2009hc,Alexander:2009tp}. We shall find it convenient
to define the parameter $\xi \equiv \alpha^{2}/(\beta \kappa)$, as CS corrections will be proportional
to it.  

If we move the C-tensor to the right-hand-side of the modified field equations, and consider this an effective stress-energy tensor, then the energy conditions are not necessarily satisfied in an arbitrary background. 
One can see that the sign of $\chi^{a} \chi^{b} G_{ab}$ depends not only on 
$\chi^{a} \chi^{b} T_{ab}^{\rm mat}$, but also on $\chi^{a} \chi^{b} T_{ab}^{\vartheta}$ and 
$\chi^{a} \chi^{b} C_{ab}$, where $\chi^{a}$ is a timelike or null vector, depending on the energy condition
considered. In certain non-trivial spacetimes, such as 
the spinning black hole solution recently found in~\cite{Yunes:2009hc,Konno:2009kg} or the perturbed 
Schwarzschild spacetime considered here, the energy conditions are not satisfied.

{\emph{Linear Stability Analysis}}. We study the modified field equations in perturbation theory, decomposing the full, spacetime metric into $g_{ab} = \bar{g}_{ab} + \epsilon_{h} \; h_{ab}$ and the 
full CS scalar field into $\vartheta = \bar{\vartheta} + \epsilon_{h} \; \delta\vartheta$. We work to first 
order in the perturbation book-keeping parameter $\epsilon_{h}$, which labels the order of the 
perturbation. The background quantities $(\bar{g}_{ab},\bar{\vartheta})$ are solutions 
to the unperturbed field equations, while $(h_{ab}, \delta\vartheta)$ are small perturbations away 
from this background.  

We consider background solutions where the metric is vacuum and where both the scalar field and the 
Pontryagin invariant vanish.  Examples of such solutions are the Minkowski and Schwarzschild 
spacetimes. We seek solutions to the perturbed metric and CS field of the form 
\ba
h_{ab} = A_{ab}(t,x^{j}) \; e^{i \phi(t,x^{k})/\epsilon_{\phi}}, 
\nonumber \\ 
\delta\vartheta = B(t,x^{j}) \; e^{i \phi(t,x^{k})/\epsilon_{\phi}}\,.
\ea
We further impose the geometric optics approximation, where we require that the phase 
$\phi$ varies much faster than the amplitudes $(A_{ab},B)$. This is enforced by requiring 
that the phase book-keeping parameter $\epsilon_{\phi} \ll 1$. With the above ansatz
we are also restricting this analysis to steady-state solutions, 
hence the CS scalar and metric perturbation have the same phase.

The modified field equations and equation of motion for the CS field can now be expanded bivariately 
in $\epsilon_{h} \ll 1$ and $\epsilon_{\phi} \ll 1$. The dominant term in the expansion is of 
${\cal{O}}(\epsilon_{h},\epsilon_{\phi}^{2})$. The perturbed equation of motion for the CS field 
becomes
\be
\label{pert-EOM}
(\delta \vartheta) k_{a} k^{a} =  -\frac{\alpha}{2 \beta} \bar{C}_{abcd} \bar{\epsilon}^{abef} 
h^{d}{}_{f} k^{c} k_{e}\,,
\ee
while the perturbed CS field equations become
\be
\label{pert-Feq}
h_{ab} k^{c} k_{c} = \frac{\alpha}{\kappa} (\delta \vartheta) \bar{\epsilon}^{def}{}_{(a} 
\bar{C}_{b)}{}^{g}{}_{ef} k_{g} k_{d}\,.
\ee
Here we have defined the four wavevector $k^{a} \equiv \nabla^{a} \phi$, and 
${{\bar C}_{abcd}}$ is the background Weyl tensor (which is equal to the background Riemann tensor
since the background is vacuum). 

The perturbed field equations and the equation of motion for the perturbed scalar field can be combined
to obtained a modified dispersion relation, namely
\ba
(k_{a} k^{a})^{2} &=& - \frac{\xi}{2} \bar{\epsilon}^{abef} \bar{\epsilon}^{hij}{}_{d} \bar{C}_{abc}{}^{d} 
\bar{C}_{f}{}^{g}{}_{ij} k^{c} k_{e} k_{g} k_{h} 
\nonumber \\
&=& 2 \xi \; ({}^{*}\bar{C}^{f}{}_{ec}{}^{d}) \; ({}^{*}\bar{C}_{hdfg}) \; k^{c} k^{e} k^{g} k^{h}\,.
\ea
For notational convenience, we define the tensors $S_{ab}$ and $W_{ab}$ in analogy with 
the magnetic and electric parts of the Weyl tensor  
\ba
W_{ac} &=& {\bar C}_{abcd} k^{b} k^{d},
\nonumber \\
S_{ac} &=& {}^{*}\bar{C}_{abcd} k^{b} k^{d}  
= \frac{1}{2} \epsilon_{cd}{}^{ef} C_{abef} k^{b} k^{d}\,.
\ea
Notice that $W_{ac}$ plays the role of the electric part of the Weyl tensor $E_{ab}$ and $S_{ac}$ that of the 
magnetic part $B_{ab}$, except that here the Weyl tensor is contracted with the four-wavevector $k^{a}$, 
instead of a timelike unit vector~\cite{Bonnor:1995zf}. In terms of these quantities, the modified dispersion 
relation can be written in the following simple form
\be 
(k_{a} k^{a})^{2} =  2 \xi \; (S^{ab}) \; (S_{ab})\,.
\label{mod-disp}
\ee

Some interesting conclusions follow from these results. First, the right-hand side of Eq.~\eqref{mod-disp}
vanishes trivially for all Petrov type-O (conformally flat) spacetimes, such as Minkowski,
simply because the Weyl tensor vanishes. Therefore, GWs in
dynamical CS gravity have the usual dispersion relation $k_{a} k^{a} = 0$ to linear order in these spacetimes
and they travel at exactly the speed of light. 
Second, when $\xi\geq 0$, which is natural as this would impart positive
energy to the scalar field, then linear stability requires that $S^{ab} \; S_{ab} \geq 0$ on the given
background. Using the results of~\cite{Bonnor:1995zf}, it is
straight foward to show that all Petrov type-N and type-III (pure gravitational wave) spacetimes have
$k_{a} k^{a} \geq 0$ to linear order. If not strictly $0$, wave speeds will differ from that of light.
For spacetimes that have less symmetries in the Weyl tensor, the right-hand side of 
Eq.~\eqref{mod-disp} does not vanish in general, and we are not aware of results regarding its positivity.
Such spacetimes would need to be examined on a case-by-case basis, as we do next for Schwarzschild.

{\emph{The Speed of Gravitational Waves in Dynamical CS Theory on a Schwarzschild Background}}.
The modified dispersion relation of Eq.~\eqref{mod-disp} can be solved given a specific background
that is a solution to CS modified gravity, and here we do so for the Schwarzschild metric. 
Parameterizing the four wave-vector as $k^{a} \equiv [\Omega,k_{1},k_{2},k_{3}]$, we find that in a 
Schwarzschild background, keeping only the correction that is lowest order in $\zeta$,
\be
\Omega  =  {\Omega _{\rm GR}} \left [ 1 \pm {\frac {3 {m^3}} { r^3}} {\zeta ^{1/2}} 
\left(1 - \frac{k_1^2}{\Omega_{\rm GR}^2 f^2}
\right)
\right ]\,.
\label{CS-omega}
\ee
Here $(t,r,\theta,\phi)$ are standard Schwarzschild coordinates, $m$ is the black hole mass, $f = 1 - 2 m/r$,
$\zeta = \xi/m^{4}$ and the GR result is 
\be
\Omega_{\rm GR} =\pm \frac{1}{f} \sqrt{k_{1}^{2} + f r^{2} k_{2}^{2} + f r^{2} \sin^{2}{\theta} k_{3}^{2}}\,.
\label{CS-omega-GR}
\ee
This solution to the CS modified dispersion relation is different from the amplitude birefringent one found 
in~\cite{Alexander:2007:gwp,Yunes:2010yf}, because here
we are considering the dynamical theory, instead of the non-dynamical one and we are
treating $\delta \vartheta$ as a perturbation, with no background $\bar{\vartheta}$.

This solution deserves further discussion. First, notice that when dealing with an ingoing or outgoing 
spherical wavefront, such that $k_{2,3} = 0$, there is no CS correction, $\Omega = \Omega_{\rm GR}$. 
This makes sense, as a spherical wave front respects the spherical symmetry of the background, not providing
the CS correction with any vector to couple to. Conversely, the maximal deviation from GR is achieved
when there is no radial part to the wave vector. Second, the correction
vanishes asymptotically and at the event horizon $r = 2 m$ ($f = 0$), 
as can be seen by substituting Eq.~\eqref{CS-omega-GR} into Eq.~\eqref{CS-omega}. Thus, gravitational
waves still ``see'' the same event horizon of the underlying geometry.
Third, the CS correction to $\Omega_{\rm GR}$ is a term of 
third post-Newtonian order, as it is proportional to $[G m/(c^{2} r)]^{3}$, momentarily reinstating physical units.
Fourth, the correction scales as $\zeta^{1/2}$, instead of $\zeta$ as found for the CS modified 
Kerr solution in~\cite{Yunes:2009hc}, implying the effect found in this paper is, in this sense, larger.

Perhaps the most interesting feature of the solution found above is the appearance of sub- and 
superluminal modes. The modified dispersion relation is a quartic equation for $\Omega$, instead 
of a quadratic one as in GR. Thus, one now finds four independent propagating modes: two representing
ingoing and outgoing waves, and for each of these, two more representing subluminal and superluminal
propagation. In fact, one can check that the characteristics are no longer null, but now satisfy
\be
k_{a} k^{a} =  \pm {\frac {6{m^3}} r} {\zeta ^{1/2}} ({k_2 ^2} + {k_3 ^2}{\sin ^2} \theta )
\,.
\ee

The most general solution is a superposition of all modes:
\ba
h_{ab} &=& A_{ab}^{o,+} e^{i \phi_{+}} + A_{ab}^{o,-} e^{i \phi_{-}} 
+ A_{ab}^{i,+} e^{-i \phi_{+}} + A_{ab}^{i,-} e^{-i \phi_{-}}\,,
\nonumber \\
\delta \vartheta  &=& B^{o,+} e^{i \phi_{+}} + B^{o,-} e^{i \phi_{-}} 
+ B^{i,+} e^{-i \phi_{+}} + B^{i,-} e^{-i \phi_{-}}\,, \nonumber\\
\ea
where the phase $\phi_{\pm}$ is subject to Eq.~\eqref{CS-omega} 
with either the plus or minus sign. In the regime where $k^a k_a \neq 0$ 
the amplitude coefficients are not independent quantities, but instead
related via Eqs.~\eqref{pert-EOM}-\eqref{pert-Feq}, which reduce to 
\be
A_{ab} = - \frac{2\alpha}{\kappa} \frac{B}{k_{c} k^{c}} S_{ab},
\qquad
B = - \frac{\alpha}{\beta} \frac{A^{ab} S_{ab}}{k_{c} k^{c}}\,.
\label{mixing}
\ee
because $S_{ab} S^{ab} = (k_{c} k^{c})^{2}/(2\xi )$ by Eq~\eqref{mod-disp}.
Part of the reason for this tight coupling between the fields
is that we are only looking at steady-state solutions.
In other words, nothing prevents us from choosing initial conditions
to the original system~\eqref{eom} that consists of an arbitrary
combination of CS scalar perturbation plus gravitational wave
perturbation. However, due to the coupling in the evolution equations,
gravitational wave perturbations will source CS scalar perturbations,
and vice-versa, and if eventually a steady state is reached, the
ratios of amplitudes will be given by Eq.~\eqref{mixing}. This implies, for example,
that the scattering of a pure gravitational wave off a Schwarzschild black hole in CS
modified gravity would source an outgoing scalar field component.

{\emph{Summary}}.~In this paper we have shown that dynamical CS gravity is linearly stable
to high-frequency perturbations on a variety of physically relevant backgrounds.
An interesting further consequence, valid on general non-trivial backgrounds, is that the
speed of gravitational waves is not equal to that of light.
One might worry that such superluminal propagation violates causality. 
Superluminal propagation of a field, however, is not necessarily related to 
causality violation. In particular, the perturbation to the metric is small,
and so the causal structure of the spacetime is essentially unaltered.
Furthermore, one cannot use the gravitational field to communicate signals
acausally---that would require arbitrarily fast speeds of propagation,
which is not the case here (see for example~\cite{Babichev:2007dw} for
a similar discussion on such issues arising in k-{\em essence}
cosmological models). 

One possible effective interpretation is to think of the background geometry as
definiing a preferred reference frame, or ``aether'', in which gravitational wave disturbances propagate,
breaking local Lorentz invariance.
In this sense, one can think of CS gravity as promoting the background spacetime 
to a non-vacuum medium on which gravitational waves propagate with an anomalous dispersion relation, 
similar to the analogous situation of electromagnetic waves propagation in an anomalous medium. 

An interesting consequence of such anomalous GW dispersion is the plausible temporal separation of
GW and electromagnetic pulses that would otherwise be coincident in GR. Our results suggest that as a
GW scatters off a non-trivial background, two modes will generically be excited (a super-luminal and a 
sub-luminal one) that will separate as these waves traverse the non-trivial background. For example, if the computed 
CS effect persists when galaxies act as the lens of cosmological GWs, then these waves might accumulate a delay.
However, a back of the envelope estimate, together with current constraints on $\xi^{1/4} < 10^{4} \;  \rm{km}$~\cite{Yunes:2009hc}, suggest 
that this effect might not currently be measurable. This is because $\zeta = \xi/m^{4} \ll 1$ when one saturates the 
current constraints on $\xi$ and uses galactic masses on the order of $10^{12} M_{\odot}$. As argued 
in~\cite{Yunes:2009hc,Sopuerta:2009iy}, the best constraints on dynamical CS gravity arise for small compact 
objects, where the Riemann tensor, which scales as the inverse of the mass squared, is large.

{\emph{Acknowledgments}}.
We are grateful to
Carlos Barcelo,
Mihalis Dafermos, 
Vitor Cardoso,
Leonardo Gualtieri,
and David Spergel, 
for useful suggestions and comments.  Some calculations used the computer algebra systems MAPLE, in combination with the GRTensorII package~\cite{grtensor}. FP and NY acknowledge support from the NSF grant PHY-0745779, and FP acknowledges support from the Alfred P. Sloan Foundation. DG acknowledges support from NSF grant PHY-0855532.
\bibliographystyle{apsrev}
\bibliography{review}

\end{document}